**Effects of local interaction and dispersal on the dynamics of size-structured populations**

Thomas Adams* (1), Graeme Ackland (1), Glenn Marion (2) and Colin Edwards (3)

(1) School of Physics and Astronomy, The University of Edinburgh, EH9 3JZ, Scotland
(2) Biomathematics and Statistics Scotland, EH9 3JZ
(3) Forest Research, Northern Research Station, Midlothian, EH25 9SY

* E-mail: tomadams@ekit.com

**Abstract**

Traditional approaches to ecosystem modelling have relied on spatially homogeneous approximations to interaction, growth and death. More recently, spatial interaction and dispersal have also been considered. While these leads to certain changes in community dynamics, their effect is sometimes fairly minimal, and demographic scenarios in which this difference is important have not been systematically investigated.

We take a simple mean-field model which simulates birth, growth and death processes, and rewrite it with spatially distributed discrete individuals. Each individual's growth and mortality is determined by a competition measure which captures the effects of neighbours in a way which retains the conceptual simplicity of a generic, analytically-solvable model. Although the model is generic, we here parameterise it using data from Caledonian Scots Pine stands. The dynamics of simulated populations, starting from a plantation lattice configuration, mirror those of well-established qualitative descriptions of natural forest stand behaviour; an analogy which assists in understanding the transition from artificial to old-growth structure.

When parameterised for Scots Pine populations, the signature of spatial processes is evident, but they do not have a large effect on first-order statistics such as density and biomass. The sensitivity of this result to variation in each individual rate parameter is investigated; distinct differences between spatial and mean-field models are seen only upon alteration of the interaction strength parameters, and in low density populations. Under the Scots Pine parameterisation, dispersal also has an effect of spatial structure, but not first-order properties. Only in more intense competitive scenarios does altering the relative scales of dispersal and interaction lead to a clear signal in first order behaviour.

**Introduction**

Ecological communities with clearly defined size and spatial structure have been studied for many decades. Much work has focussed on the outcome of competitive interactions between species (Tilman & Wedin, 1991; Chesson, 2000; Perry et al., 2003) and in the analysis of both size-structured (see e.g. Sinko & Streifer, 1967) and, more recently, spatially-structured population models (see e.g. Bolker & Pacala, 1997; Law et al., 2003). However, a mechanistic understanding of the dynamics of real communities, structured in both size and space, has been limited by a lack of application of simple models, amenable to analysis and approximation, to the communities in question (Gratzer et al., 2004).

In the case of forest populations, a practical understanding of the general patterns and forms observed in population dynamics is well established (Franklin et al., 2002). A great body of simulation models for multi-species communities (e.g. Botkin et al., 1972; Pacala et al., 1996; Busing & Mailly, 2004) has also been developed over the years. However, the potential for useful results derived from the study of monocultures is far from exhausted. While pure forest monocultures may be rare in nature, many communities are dominated by a single species, and their theoretical study presents a clear and accessible way to understand and identify driving processes and mechanistic changes over time, and the effect of demographic rates on fundamental properties such as size and spatial structure (Bolker & Pacala, 1997, Law et al., 2003).

An important concept in forest conservation is that of an "old-growth" state. This is an autogenic state which is obtained through an extended period of growth, mortality and regeneration, in the absence of external disturbances, which may take several centuries to attain (Oliver & Larson, 1996). It is often seen as an "equilibrium" state, and is characterised by a fully represented (high variance) age and size structure, and non-regular spatial pattern. The habitat created in this state is often considered a paradigm of what conservation oriented forest management might hope to achieve (Schutz, 2002); as such, we would like to understand more clearly the processes that affect its general properties. Here, we develop and directly apply a generic process-based model, closely related to those of Bolker & Pacala (1997) and Law et al. (2003), to understanding the key elements of community behaviour, from a plantation through to old-growth. Data for Scots Pine (*Pinus Sylvestris* L.) populations at various stages of development provide a baseline for comparison. However, our also investigation focuses more generically on the importance of spatial effects in such populations, and identifies demographic scenarios in which their inclusion is essential.

As a null hypothesis, a generic size-structured "mean-field" model is introduced in the following section. This is unable to explain all qualitative aspects of community behaviour, and so an analogous spatial individual based model is also presented and parameterised. An initial growth dominated period gives way to a reduction in density and a meta-stable state governed by reproduction and mortality, all of which correspond with field observations of the growth of communities of a range of species. While spatial interactions and dispersal do alter the dynamics of our model populations, the impact on first-order properties using the Scots Pine parameterisation is fairly small. We identify the regions of parameter space in which spatial interactions would become more important, enabling an assessment of the scenarios in which a mean-field model is likely to be acceptable for the representation of size-structured populations. The approach of gradually extending a simple "null" simulation model, and comparing output with diverse population data, allows clearer identification and understanding of the drivers of ecosystem dynamics and the steady state.

**Materials and Methods**

MEAN-FIELD MODEL

Consider a population of individuals, each characterised by a single size measure s. This may be mass, height or any other metric, but in the case of trees is usually taken to be "diameter at breast height", or dbh. In the mean-field case all individuals have an identical experience, and we are thus interested in the evolution of the density of individuals across the range of possible sizes, n(s,t).

We use the Gompertz model for individual growth, reduced by competitive interactions (Wensel et al., 1987). This function has been applied successfully to both trees and other plants (Zeide, 1993; Schneider et al., 2006). Of Richards (1959) type asymptotic growth models, it was found to be the best fitting descriptor of growth in statistical analysis of individual tree growth increment data, accounting for the effects of interaction (results not shown). The growth rate is

$$G(s,t) = \frac{ds(t)}{dt} = s(t)\left(\alpha - \beta \ln(s(t)) - \gamma \Phi(s,t)\right) \quad (1)$$

where $s$ is the size of an individual, $\Phi$ is the competition experienced at that size (dependent upon current population state) and $\alpha$, $\beta$, $\gamma$ are species dependent parameters. This leads to an asymptotic maximum size, $s^* = \exp(\alpha/\beta)$ if competition is absent. Under intense competition, the right hand side of Equation 1 may be negative. Following Weiner et al. (2001), we fix $G(s,t)=0$ in this case.

Competition is assumed to be asymmetric, and takes a form which depends on the density, size and relative size of the other individuals in the population,

$$\Phi(s,t) = \int_{s'} n(s',t) f(s,s') ds' \quad (2)$$

$$f(s,s') = s'\left(\tanh\left(k_s \ln\left(\tfrac{s'}{s}\right)\right) + 1\right) \quad (3)$$

The tanh function allows anything from symmetric ($k_s=0$) to completely asymmetric competition

($k_s \to \infty$) (Schneider et al., 2006). Multiplying interaction by the size $s'$ of the neighbour considered reflects the increased competition between larger individuals, independent of the size difference (consider two tiny individuals with given separation/size-difference, compared to two large ones with the same separation/difference).

Mortality occurs at a rate

$$M(s,t) = \mu_1 + \mu_2 \Phi(s,t) \tag{4}$$

$\mu_1$ is a fixed baseline (Wunder et al., 2006), and $\mu_2$ causes individuals under intense competition to have an elevated mortality rate (Taylor & MacLean, 2007).

Finally, the boundary condition for the process is given by the establishment of seedlings. Existing trees thus produce offspring at a rate determined by their basal area (Strigul et al., 2008). The population's rate of seed production is

$$B(t) = f \int_s n(s,t) \frac{\pi s^2}{4} ds \tag{5}$$

where f is the birth rate per m$^2$ basal area. The fecundity of trees and accurate quantification of seed establishment success is a long standing problem, due to the combination of seed production, dispersal, neighbourhood and environmental effects involved (Clark et al., 2004; Gratzer et al., 2004). Sub-models for regeneration are often used (e.g. Pacala et al., 1996), but for simplicity we remove this stage of the life cycle from the model by considering only individuals above a minimum size of 1cm dbh. We assume that an individual takes y years to reach this size, and thus define a probability of seed establishment/survival: $P_e(t) = \left(1 - (\mu_1 + \mu_2 \Phi(1,t))\right)^y$. This ignores fluctuations in population state throughout the establishment period, but should be a good approximation in the steady state.

The evolution of the size density distribution n(s,t) is thus described by the equation

138 $$\frac{\partial n(s,t)}{\partial t} + \frac{\partial (n(s,t)G(s,t))}{\partial s} = -M(s,t)n(s,t) \quad (6)$$

139 with boundary condition $n(1,t)=B(t)P_e(t)$. This dynamical model is similar to that discussed by

140 Sinko & Streifer (1967) and Angulo & Lopez-Marcos (2000), but additionally incorporates a

141 population state dependent interaction effect in the functions G(s,t) and M(s,t).

142

143 EQUIVALENT SPATIAL MODEL

144

145 The model can be readily generalised to a Markovian stochastic birth-death-growth process in

146 continuous (two-dimensional) space. Individuals i=1,..,N are characterized by position and size,

147 which jointly define the state space of the process.

148

149 Interaction (and hence growth, mortality and establishment) are not not strictly governed by size as

150 they are in the mean-field model, since now neighbourhood varies across individuals. To generalise

151 the model to include spatial dependence, we rewrite the competition as

152 $$\Phi_i(t) = \sum_{j \in \omega_i} f\left(s_i(t), s_j(t)\right) g\left(\vec{x}_i, \vec{x}_j\right) \quad (7)$$

153 where $\omega_i$ is the set of all individuals excluding i. $s_i$ is the size of tree i and $\vec{x}_i$ its position. Note that

154 this is a sum over individuals, as opposed to the integral over the density function in Equation 2.

155

156 Following Raghib-Moreno (2006); Schneider et al. (2006), the spatial component of interaction is

157 introduced with a Gaussian function of distance to neighbours

158 $$g\left(\vec{x}_i, \vec{x}_j\right) = \frac{k_d^2}{\pi} \exp\left(-k_d^2 \left|\vec{x}_i - \vec{x}_j\right|^2\right) \quad (8)$$

159 where $k_d$ defines the decay of interaction with separation. Individual growth and mortality rates

160 vary accordingly, by direct replacement of the interaction function.

161

The level of competition now varies between individuals of the same size, depending upon their spatial location in relation to others. Consequently, Equations 1 and 4 (respectively, the growth rate and mortality rate at a given size in the mean-field model) must instead be defined for each individual in the population. That is

$$G_i(t) = s_i(t)\left(\alpha - \beta \ln(s_i(t)) - \gamma \Phi_i(t)\right) \qquad (9)$$

$$M_i(t) = \mu_1 + \mu_2 \Phi_i(t) \qquad (10)$$

Reproduction is also computed on an individual basis (that is, $B_i(t) = \frac{f \pi s_i^2}{4}$). Dispersal of offspring from parents is considered in two generic forms: either randomly (with equal probability to any location in the population arena), or drawn from a Gaussian distribution (a dispersal kernel – as Equation 8 but with parameter $k_b$). Establishment uses $P_e(t)$ as above, using the spatial interaction function (7) in place of (2).

The mean-field model (Equation 6) may be derived directly from a differential equation approximating the spatial model described above, making the assumption that the pair density of individuals with sizes s and s' separated by distance r, n(s,s',r), can be approximated as n(s)n(s') (manuscript in preparation).

STATISTICS AND SIMULATION

Community structure is tracked using various metrics: density (number of individuals per m²), total basal area (mean field: $\int_s n(s)\left(\frac{\pi s^2}{4}\right) ds$, spatial: $\sum_i \frac{\pi s_i^2}{4}$), size and age density distributions, and pair correlation (PCF) and mark correlation (MCF) functions (relative density and size multiple of pairs at given separation, Penttinen et al., 1992; Law et al., 2009). All presented spatial model results presented have mean and standard deviation (in figures, lines within grey envelopes), which are computed from 50 repeat simulation runs. The simulation arena represents a 1ha plot (100x100m).

187  Periodic boundary conditions are used to remove edge effects. Due to the scale of the kernels used,
188  results are not significantly altered by increasing arena size.

189

190  The mean-field model is integrated using an explicit forward-difference numerical scheme, with a
191  size step of 0.1cm and a time step of 0.2 years. The spatial model is integrated numerically in
192  continuous time by means of the Gillespie algorithm (Cox & Miller, 1965; Gillespie, 1977); this
193  generates a series of events (i.e. growths, births, deaths) and inter-event times. After any given
194  event, the rate ($r_{event}$) of every possible event that could occur next is computed. The time to the next
195  event is drawn from an exponential distribution with rate $R = \sum r_{event} = \sum_i \left( B_i(t) + G_i(t) + M_i(t) \right)$; the
196  probability of a particular event occurring is $r_{event}/R$ .

197

198  PARAMETERISATION FOR CALEDONIAN SCOTS PINE

199

200  We use data from two broad stand types (collected in Scotland by Forest Research, UK Forestry
201  Commission): plantation and "semi-natural" (see Edwards & Mason, 2006; Mason et al., 2007).
202  Plantation datasets (6x1.0ha stands) from Glenmore (Highland, Scotland) incorporate location and
203  size, allowing comparison of basic statistics at a single point in time (stand age approximately 80
204  years). Semi-natural data is available from several sources. Spatial point pattern and increment core
205  data (measurements of annual diameter growth over the lifespan of each tree, at 1.0m height) for
206  four 0.8ha stands in the Black Wood of Rannoch (Perth and Kinross, Scotland) allows estimation of
207  growth and interaction parameters. Location and size measurements (in 1997) from a 1.0ha semi-
208  natural stand in Glen Affric (Highland, Scotland) provide another basis for later comparison.

209

210  Our simulations use a dispersal kernel with identical spatial scale to the interaction kernel, and an
211  establishment time (*y*) of 20 years, in accordance with field studies of Scots Pine regeneration
212  (Sarah Taylor, unpublished data). In none of the stands is there adequate information to reliably

213   estimate mortality ($\mu_1$, $\mu_2$) or fecundity (*f*). These are thus tuned to meet plantation and steady state

214   (semi-natural stand) density. The baseline mortality rate used gives an expected lifespan of 250

215   years (Featherstone, 1998; Forestry Commission, 2009).

216

217   A nonlinear mixed effects (NLME) approach (Lindstrom & Bates, 1990) was used to estimate

218   growth parameters α, β, and γ. Best-fitting growth curves were computed for each of a subset of

219   individuals from two of the Rannoch plots, and the mean, standard deviation and correlation

220   between each parameter within the population was estimated. Details are given in Appendix S1.

221   Mean values for α and β are used for simulation, though large variation between individuals was

222   observed. γ was difficult to estimate from the semi-natural data, its standard deviation being larger

223   than its mean; a consequence of the fact that interaction does not explain a majority of variation in

224   individual growth (see Appendix S2). However, it has a large effect on the simulated "plantation"

225   size distribution (Appendix S3). Therefore, a value slightly lower than the estimated mean was used

226   in order to better match the size distribution in both plantation and semi-natural stages.

227

228   $k_d$ was selected to provide an interaction neighbourhood similar to previous authors (e.g. Canham et

229   al., 2004). $k_s$ determines early plantation size distribution, and was selected accordingly; it has

230   minimal effect on long-run behaviour. Parameter values used for model Scots Pine populations are

231   shown in Table 1. Sensitivity to parameter variation over broad intervals was also tested (Appendix

232   S3).

233

234   A standard planting regime implemented in Scots Pine plantations is a 2m square lattice, typically

235   on previously planted ground. Old stumps and furrows prevent a perfectly regular structure being

236   created, so our initial condition has 1cm dbh trees with small random deviations from exact 2m

237   square lattice sites, which more closely resembles observed planting positions. With such tuning, it

238   is found that the model is able to replicate key patterns observed in both plantation and semi-natural

239  data stands (see Appendix S1). The generic aspects of model behaviour, and specific differences

240  between its behaviour and that observed in real forests, are outlined below.

241

242  **Results**

243

244  QUALITATIVE MODEL BEHAVIOUR

245

246  Various qualitative models of forest stand development are discussed by Franklin et al. (2002), and

247  the general patterns described by these are observed in our model. Starting from the plantation

248  configuration, the model population passes through several stages (an overview of which is given

249  by Figure 1): (i) an initial growth dominated period, during which the plantation structure largely

250  remains, and the canopy closes; (ii) a period of high density-dependent mortality as the impact of

251  interactions begins to be felt; (iii) gap creation together with an increase in regeneration; (iv) the

252  long-run meta-stable state, during which stand structure is more irregular and determined by the

253  levels of mortality and birth.

254

255  The plantation structure initiated by forest management has a higher density than a natural self-

256  regenerating forest. Initially, reproduction is low, due to individuals' small size. Rapid growth of

257  the immature trees means that basal area increases rapidly (see Fig. 2a). Individual density falls

258  equally quickly due to high levels of density-dependent mortality. Stochastic variation in growth

259  and asymmetric competition lead to a gradual spread of sizes of individuals (the initial size

260  distribution is a delta peak at $s = 1cm$). Size asymmetry is often cited as a key driving force in plant

261  community dynamics (Adams et al., 2007; Perry et al., 2003; Weiner et al., 2001). In our model,

262  competitive size asymmetry is the primary factor affecting the variance (spread) of the size

263  distribution during the early stages of stand development: it is almost independent of any other

264  parameter, or even starting spatial configuration (see Table 2, Appendix S2). In the spatial model,

low reproduction means that spatial structure is governed by the starting configuration. The pair correlation function (PCF), giving the relative density of pairs of individuals with given separation (Penttinen et al., 1992), clearly shows the signature of the lattice during this stage (Fig. 2c, 80 years - peaks are at multiples of the lattice spacing). The mark correlation function (MCF) measures the relative size of individuals forming pairs at a given separation, compared to the global average (Penttinen et al., 1992), but does not provide a great deal of useful information at this stage due to the regular pattern of trees. The period described above contains the cohort establishment, canopy closure and biomass accumulation stages of Franklin et al. (2002).

The high basal area (and high competition) state generated during the "plantation" stage means that individual growth becomes stunted, and mortality rates are elevated. Basal area thus reaches a peak. Density-dependent mortality remains high, but is overtaken by density-independent (intrinsic) mortality, which opens gaps in the canopy. Consequently, more substantial regeneration begins to occur (gaps increase $P_e$ for many of the potential offspring, while high basal area ensures a large seed source) and a much broader age/size structure begins to develop. The initial regular spatial structure is erased during this period, through mortality, regeneration and differential growth. This change is apparent in both spatial correlation functions (not shown), and in maps of the stand at 300 years (Fig. 1). During this period, a real stand would also see the accumulation of woody debris (in large part arising through heightened mortality seen in our model). This is the maturation stage of Franklin et al. (2002).

In many real populations the generic properties of the observed state are substantially determined by external disturbances (and the relationship between their extent and frequency), as opposed to demographic properties alone (Turner et. al., 1993), bringing into question the utility of the terms "old-growth" or "equilibrium" in describing natural systems. Indeed, Oliver & Larson (1996) point out that due to external catastrophic disturbances, true old growth is rarely reached by many temperate forest communities. In the long run (and in the absence of external disturbance), the model reaches a steady state where fecundity, mortality and growth are in balance. Figure 2b (dotted lines) shows the typical size structure present in the long run. Only a small proportion of juveniles attain canopy size, but individuals of all sizes are present, and the asymptotic nature of growth means that individuals accumulate in the higher size classes as the system approaches equilibrium, where the size distribution stabilises. This is a consequence of the ability of trees to survive during periods when they are not growing. Caledonian Scots Pine does not readily establish in low light conditions, and consequently produces a fairly low density, open forest. Growth is also very much limited by shading from other trees, but in may cases old stunted trees are observed in Scottish stands (the implication being that shading affects growth more than it does survivorship). In the model, local reductions in canopy density thus allow trees that have stunted growth to increase in size, refilling gaps.

THE EFFECT OF SPACE

Space has been noted as having a sometimes subtle but important impact on population dynamics (Law and Dieckmann, 2000; Pacala et al., 1996). Our Scots Pine data stands demonstrate generic features of the spatial structure induced by natural processes: (i) a suppressed MCF at short ranges, and (ii) a heightened PCF at short ranges. The mean-field model cannot replicate either feature, while in the spatial model such spatial structure can be produced by local interaction and dispersal.

Local dispersal of seedlings leads to an increased PCF at short ranges (Fig. 2c), while the MCF is somewhat reduced at short ranges, due to the effect of interaction on growth. If dispersal is long-ranged (random) both size and frequency of adjacent pairs is lowered, leading to reduced PCF and MCF (see below). Computing the cross-correlation function of juveniles and mature trees for the semi-natural stands shows that indeed there is either zero or negative correlation between their locations (not shown). Some authors (e.g. Barbeito et al., 2008) have noted that regeneration sometimes occurs in explicitly clustered patterns, and that this is not necessarily a consequence of local dispersal. The apparent contradiction between the model's steady state spatial correlation functions and the data suggests that the clustering seen in the data stands is partly due to management history, or environmental heterogeneity. In reality, spatial structure is also generated by disturbance (for example due to treefall during mortality).

Both mean-field and spatial models produce a bimodal size distribution, with peaks at the smallest size (juveniles) and just below $s^*=\exp(\alpha/\beta)$ ("canopy" individuals). However, "individuals" in the mean-field model experience competition based solely upon their size. This leads to a sharply peaked canopy density in the size distribution, as the entire population has an identical asymptotic size at the steady state. In the spatial model, the variation in competition over space leads to a blurring in size of the canopy, represented by a lower density, higher variance peak. Although explicit variation in asymptotic size is also likely to be a factor, space appears to play an important role in recreating the variability in canopy size that we see in real communities (but see also the Discussion). However, under the parameterisation shown in Table 1, the effect of space on individual density and basal area (a surrogate for population biomass) is fairly minor – the trajectories of density for mean-field and spatial models are almost indiscernible (Fig. 2a), while basal area at equilibrium is around 10% lower in the mean-field model. Such a limited impact is a common observation in temperate forest ecology (Deutschman et al., 1999, Busing and Mailly, 2004). Under what circumstances do spatial effects become more important?

It might be expected that in dense spatially interacting populations, local variation in neighbourhood would allow increased growth in comparison with mean-field interactions. However, this is not seen in our model (under either low mortality or high fecundity, Fig. 3a,b). Rather, in low density populations, the difference between the two models increases (with density/basal area in the mean-field model being comparatively higher, Fig. 3a,b) – an effect of finite area. Spatial interactions only directly affect the realised density when the overall effect of interaction is relatively strong in relation to basic population rates (the last term in each of Equations 1, 4, 9 and 10 is large). That is to say, increasing γ (the effect of interaction upon growth) or $\mu_2$ (the effect of interaction upon mortality) both widen the gap between simulated spatial and mean-field populations (mean-field populations having the lower density/basal area – Fig. 3d,e). Increasing $k_d$ (localisation of interaction in the spatial model) reduces the effective neighbourhood size and as a consequence leads to an increase in density and basal area (not shown). The only case in which our mean-field simulations produce a higher density and basal area than spatial simulations is when greatly increased strength of competitive interactions (γ, $\mu_2$) are combined with relatively short range dispersal ($k_b > k_d$, not shown).

We also investigated the impact that the dispersal kernel has on stand dynamics. Bolker and Pacala (1999) found that species' relative scale of dispersal affects their ability to invade one another. In our single species "Scots pine" populations altering the scale (distance $1/k_b$) of dispersal relative to the interaction kernel affects the spatial structure of the population (increasing $k_b$ producing a more clustered pattern), but does not affect the resulting population density as the effect of interaction is too weak (Fig. 4a). In more competitive populations (for example, increasing $\mu_2$ by one order of magnitude – Fig. 4b), longer range dispersal has a qualitatively similar, but more pronounced effect on spatial structure. It also allows offspring to escape the shade of their parents, and consequently increases both individual density and stand basal area (as found by Bolker and Pacala, 1997).

**Discussion**

Both mean-field and spatial models are in qualitative agreement with real communities, showing the same generic behaviour as the forest matures. However, the same Caledonian Scots Pine parameterisation results in a 10% lower basal area in the mean-field model compared with the fully spatial process, due to the lack of variation in competitive neighbourhood. In this case (by virtue of the parameterisation) the effect of including spatial heterogeneity is relatively weak. However, numerical exploration demonstrates that for highly competitive populations (or those in which interaction is very localised), the explicit treatment of space has a much larger effect on computed density or biomass. Increasing $\gamma$, $\mu_2$ or $k_d$ for example, all widen the gap between the mean-field and spatial model behaviour. In all cases the density and biomass of spatially interacting populations was higher.

The structure of simulated and real forests is strongly dependent on the initial conditions, even after hundreds of years. The long-time equilibrium state of the model has rather low density, with a highly varied size (diameter) distribution which appears to produce a stable canopy, with no evidence of cyclical variation in structural characteristics. The inclusion of a non-random dispersal kernel recreates the clustered pattern seen in data stands, and at the level of interaction present in Scots Pine stands does not greatly affect density or basal area (which it would do in more dense/competitive populations). However, it also weakens the signal of inhibition in the MCF, due to an increase in the number of parent offspring pairs at close separations. This discrepancy with data, and significant differences between real stands, suggest environmental (e.g. Gravel et al., 2008; John et al., 2007) or management influences. Plant/tree establishment has traditionally proven difficult to quantify accurately (Clark et al., 2004), and is certainly deserving of further work.

The model's interaction-limited growth is consistent with field observations. However, in model parameterisation and tuning, we found significant variation in growth trajectories between individual trees, which is impossible to explain by recourse to interaction (even when this is allowed to accumulate over time – result not shown). While the basic growth, birth and death parameters could be taken as constant for all trees, it proved necessary for the maximum size

(determined by β) to be drawn from a distribution. This may represent either genetic diversity (Provan et al., 1998) or a variation in the ability of a given location to support a tree, but we do not have relevant data for the stands in question. The robustness of model behaviour to the inclusion of such variation (see Appendix S2) suggests that the generic results that we have obtained should generalise to multi-species communities (provided that the questions being asked relate to bulk properties such as basal area, as opposed to species composition, for example).

In modelling complex real-world systems in ecology a common approach is to develop detailed application specific models (Botkin et al., 1972; Busing & Mailly, 2004). While this can be successful, such models are often difficult to parameterise given the available data, and by their nature tend to focus on system-specific features. In contrast, generic models are of great interest to theoretical ecology because they facilitate understanding of common or universal properties of ecosystems (Bolker et al., 2003; Law et al., 2009; Weiner et al., 2001). Here we have shown that generic dynamic models enable investigation of the importance of different factors and components of the life-history of a target species on population dynamics, and can be informed by empirical models. Moreover, simple generic models may also have practical advantages when applied to specific systems because they typically require relatively few parameters and, with sparse data, are less prone to over-fitting than complex models. It is inevitable that such simple models (and indeed any model) will not capture every aspect of real world systems. However, often much of this additional variability can be represented via stochasticity, implicit or explicit spatial heterogeneity and intra-individual variation in parameters.

The speed of approach to an equilibrium state is affected by disturbances (which were not implemented in this manuscript). If these are regular and major, a persistent low density state will prevail. However, small scale disturbance can benefit a stand by encouraging heterogeneity in size (through the light environment), and more rapid development of an uneven-aged structure (through regeneration) (manuscript in preparation).

CONCLUSIONS

Here we applied generic models of reproduction, competition, growth and mortality (Bolker & Pacala, 1997; Sinko & Streifer, 1967; Law et al., 2004) to real single-species population dynamics using detailed and long-term data on Caledonian Scots Pine stands. This approach was able to reproduce known qualitative and measured quantitative features of the transition from plantation to old growth stands. For such stands we found that the inclusion of explicit spatial interactions did not explain a majority of individual variation in growth, and furthermore did not have a profound effect on overall density with respect to a mean-field model. By consideration of a much wider parameter space, however, the model allowed useful generic insights into the importance of the explicit treatment of space in size-structured models of population dynamics. That they do not greatly affect the overall density and biomass of a typical modelled temperate forest population suggests that the application of mean-field models (or better, those accounting for space implicitly, e.g. Purves et al, 2008) to global issues, such as carbon and nutrient cycling, may be appropriate. However, stand level models such as ours are important for many smaller-scale goals; while our emphasis has been on understanding the mechanisms and dynamics of a population in its approach to a steady state, the model can also be used to investigate management strategies, covering diverse goals such as plantation transformation, conservation, or maximum production of timber. In conclusion, we hope that this study will prompt renewed theoretical and applied interest in dynamic models of populations structured in size and space.

**Acknowledgements**

We would like to acknowledge support from the Scottish Government and the EPSRC funded

NANIA network (grants GRT11777 and GRT11753). This work made use of resources provided by the Edinburgh Compute and Data Facility (ECDF) (http://www.ecdf.ed.ac.uk/). The ECDF is partially supported by the eDIKT initiative (http://www.edikt.org.uk).

608

609 Zeide, B. 1993. Analysis of Growth Equations. - Forest Science 39: 594-616.

610 Figure 1: Pictorial representation of 1ha Scots Pine forest. Field data (Highland, Scotland, data from
611 Forest Research, left column): 78 year old plantation in Glenmore, semi-natural stand in Glen
612 Affric. Simulated data (centre and right columns) at 50, 150, 300 and 1000 years from planting. The
613 diameter of each circle is proportional to the size (dbh) of the tree.

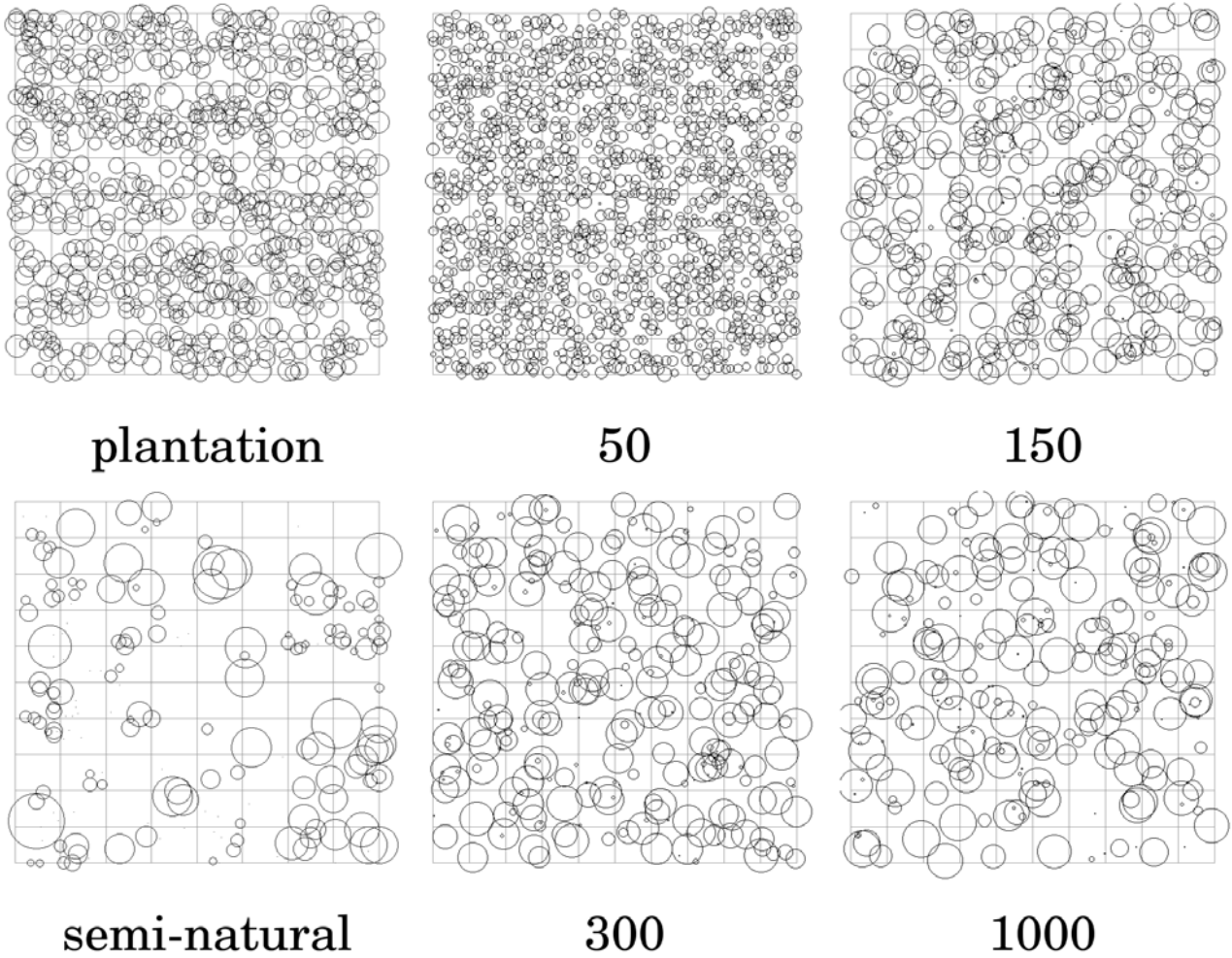

Figure 2: The transition from plantation to steady state: development of key metrics through time, based on parameters in Table 1, and including local dispersal with the same scale/parameter as the spatial interaction. Mean simulation results are represented by thin lines within a grey envelope (standard deviation), while mean-field model results are shown with thick lines. (a) Evolution of density (dashed) and stand basal area (solid line), averaged over 50 simulations of a 1ha plot. (b) Size distribution at 80 (dash-dot) and 800 (solid) years. (c) Pair correlation function - time/line style as (b), no mean-field results. (d) Mark correlation function - time/line style as (b), no mean-field results. The mean-field model produces a sharp "canopy" peak in size, whilst the spatial model has a higher variance in this region. This is more in keeping with patterns observed in real data for forest trees. The steady state stand density and basal area are not greatly affected (around 10%) by the use of a mean-field model under this parameterisation.

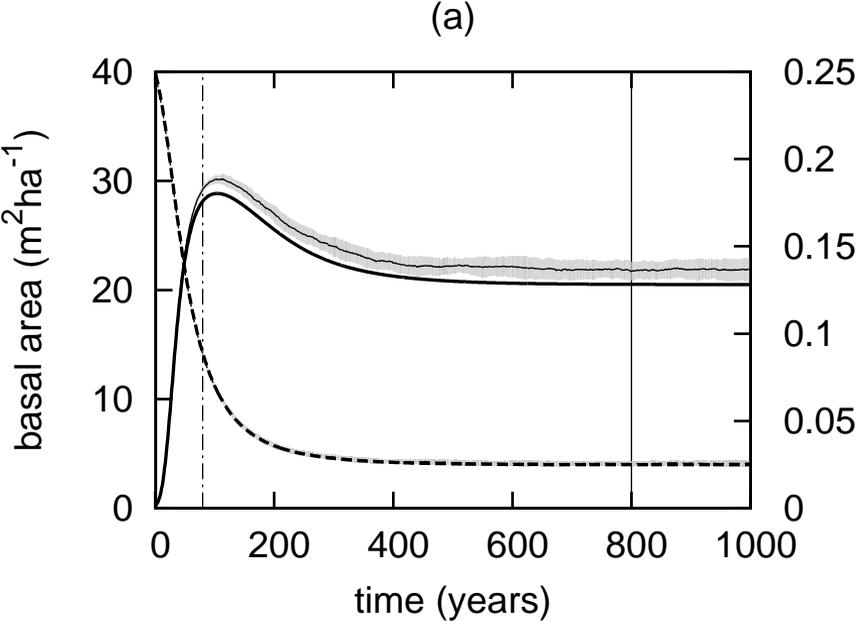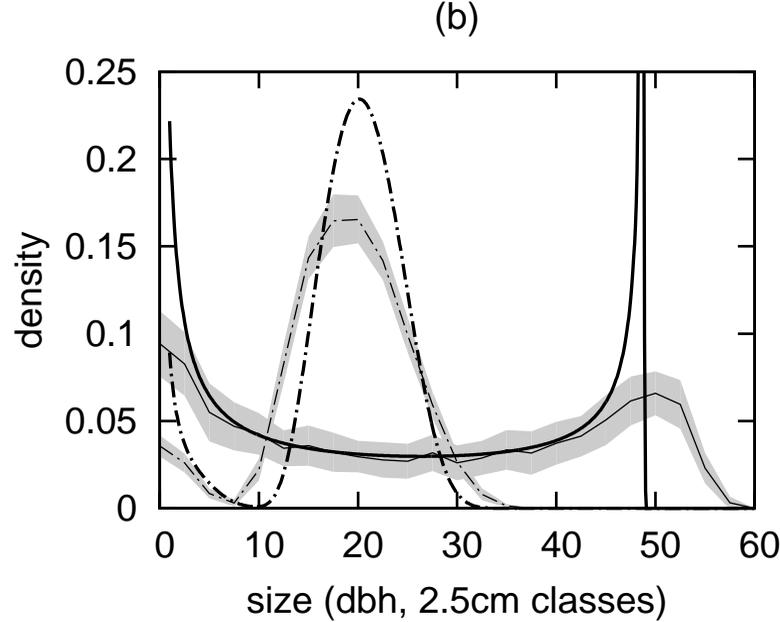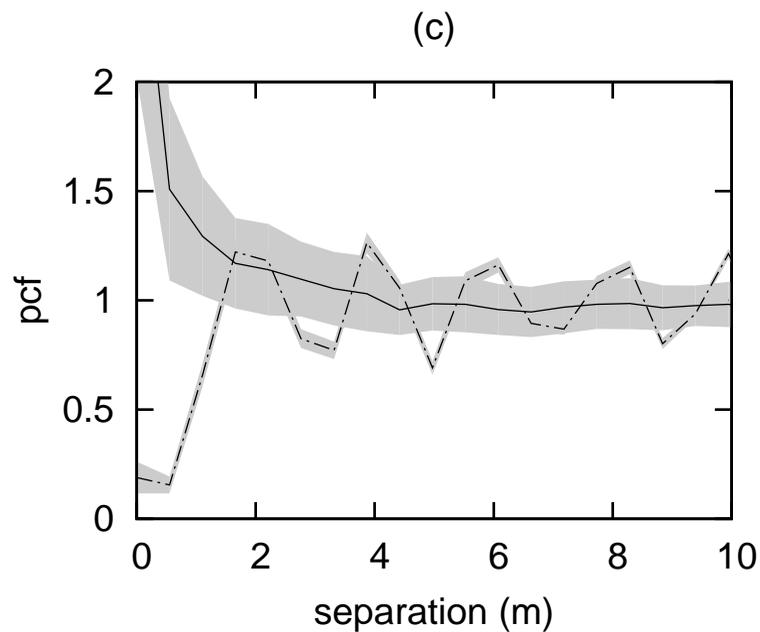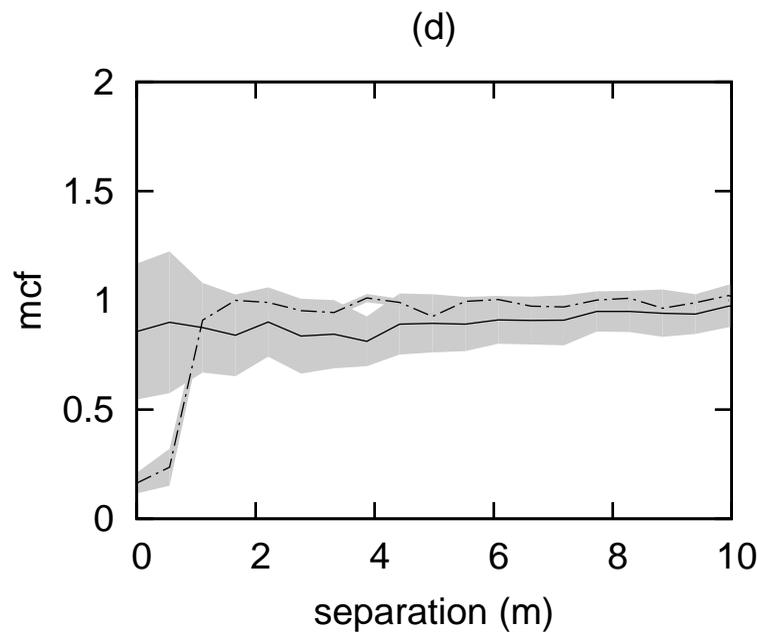

627  Figure 3: The effect of space – Sensitivity of the discrepancy between mean-field and spatial model
628  density (dashed line) and basal area (solid line) to model parameters (keeping all others equal). The
629  thick vertical line in each panel shows the "Scots Pine" parameter value (Table 1). Altering growth
630  entails altering both $\alpha$ and $\beta$, fixing their ratio (the value of $\alpha$ is shown). A mortality change entails
631  altering both $\mu_1$ and $\mu_2$ fixing their ratio (the value of $\mu_1$ is shown). The most significant differences
632  are seen on alteration of interaction parameters $\mu_2$ and $\gamma$.
633

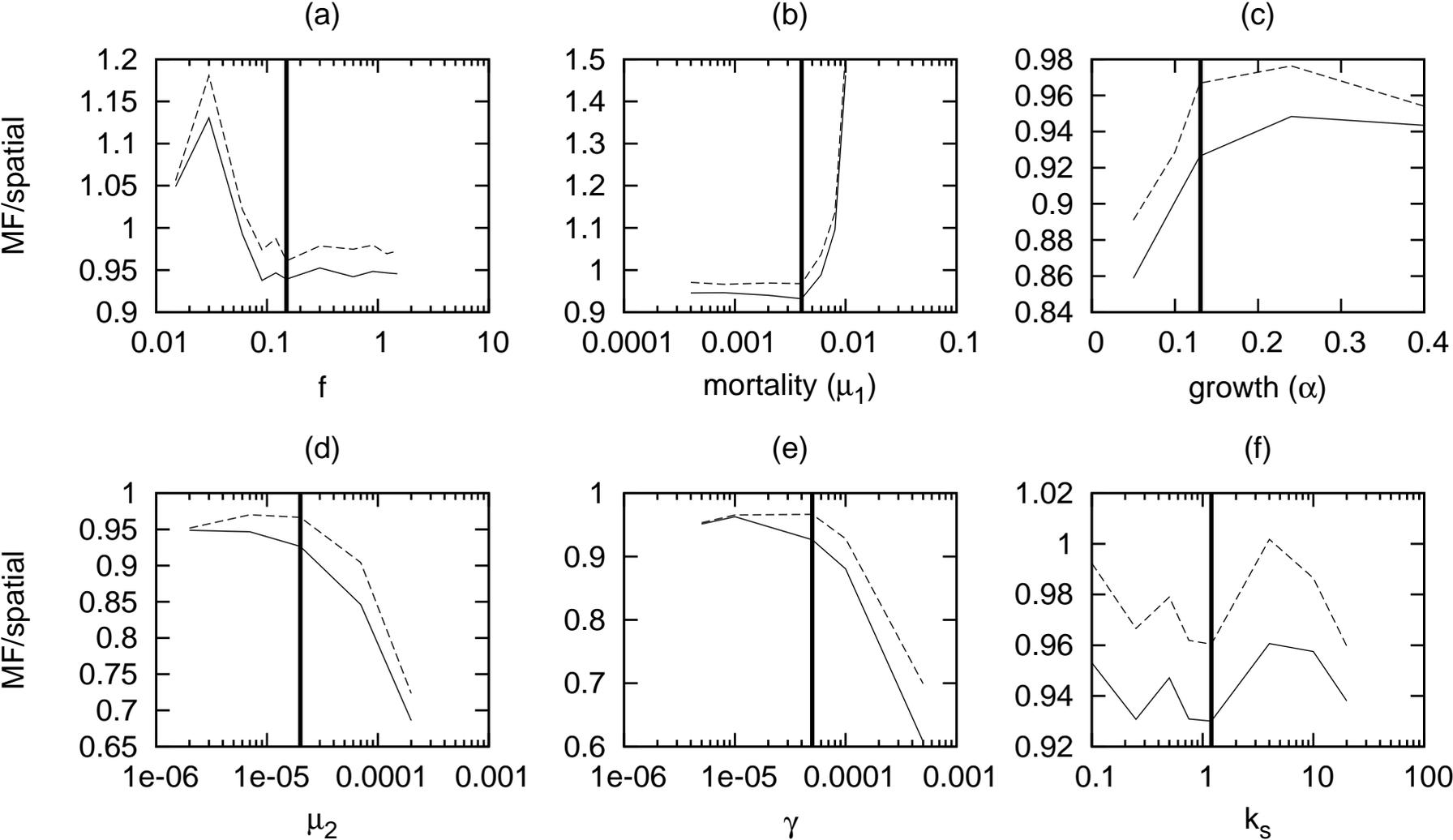

634  Figure 4: The relative scale of dispersal – altering $k_b$ (random dispersal, solid; $k_b$=0.1, dashed;

635  $k_b$=0.2, dotted) while fixing $k_d$=0.1. Insets show the change Δ in density (dashed) and basal area

636  (solid) as a result of changing $k_b$ ($k_d/k_b$ small = relatively short range dispersal). (a) Behaviour at

637  the "Scots Pine" parameterisation – spatial structure changes, but density and basal area do not. (b)

638  With stronger interaction ($\mu_2$ = 0.0002), spatial structure changes more dramatically, and

639  density/basal area also increase as dispersal becomes more global.

640

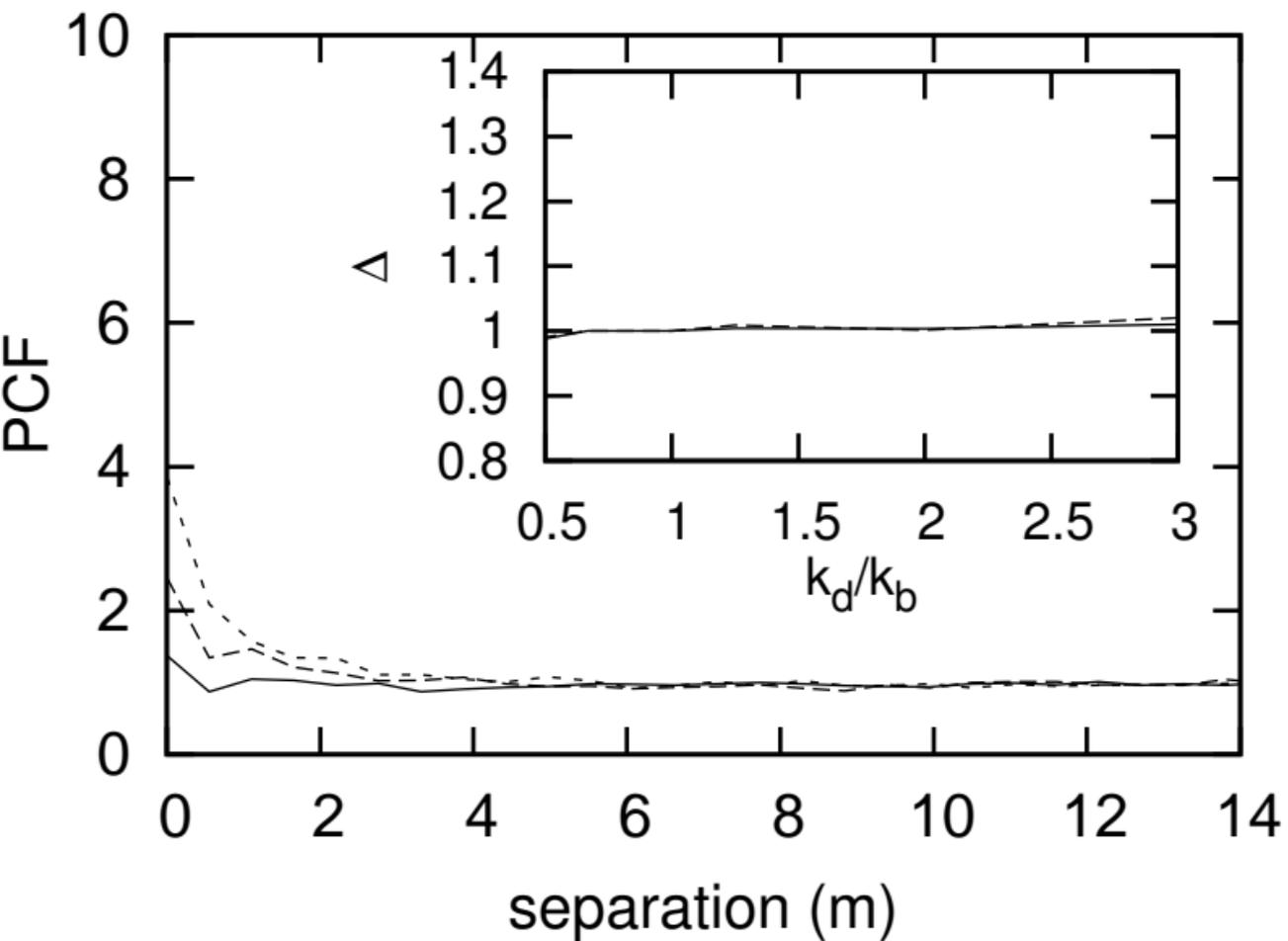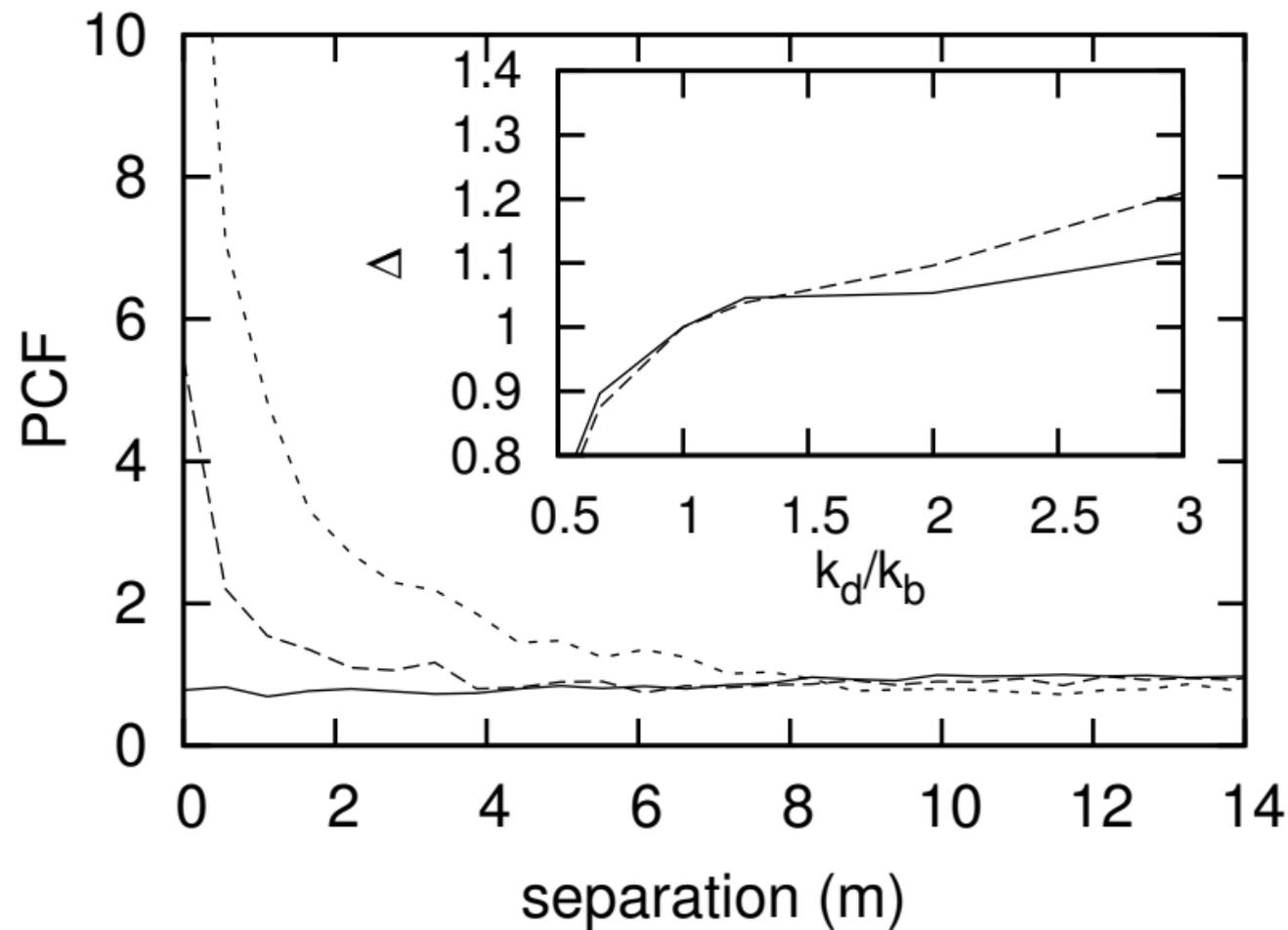

641    Table 1: Model parameters, description and values (* spatial model only).

| Parameter | Description | Value |
|---|---|---|
| **rates** | | |
| $f$ | reproduction per m$^2$ basal area | 0.15 |
| $\mu_1$ | baseline mortality | 0.004 |
| $\alpha$ | gompertz a | 0.131 |
| $\beta$ | gompertz b | 0.0316 |
| **interaction** | | |
| $\mu_2$ | mortality interaction | 0.00628 |
| $\gamma$ | growth interaction | 0.0157 |
| $k_s$ | size asymmetry | 1.2 |
| $k_d$* | distance decay | 0.1 |
| **dispersal** | | |
| $k_b$* | dispersal distance decay | 0.1 |

642

**Appendix 1 - Growth parameter estimates**

Growth parameters were estimated from increment core data (radial sections providing measurements of annual diameter growth over the lifespan of each tree, taken at 1.0m height) from several seminatural Scots Pine stands in the Black Wood of Rannoch. Parameters were estimated from individual data taken from plots ``4'' and ``6'' (5 and 7 have less well known management history). To ensure estimation based upon known competitive neighbourhoods, those individuals less than 10m from the plot boundary were excluded. Furthermore, only increments applying to growth after 1918 were used, this being the date after which management (and consequently the state of the community) is known with sufficient accuracy.

NLS is a nonlinear least squares fitting tool in R (R Development Core Team, 2005), here applied to the complete set of increment measurements. The fit computed is equivalent to assuming a single growth curve generated all data points, which are regarded as independent. NLME is another tool in R, computing a nonlinear mixed effects model (Pinheiro et al., 2009). This approach goes a step further, in computing a NLS fit for each individual in the population separately (that is, hypothesised individual growth curves). This explicitly estimates the variability present in the population by computing the mean (the "fixed effect") and standard deviation (the "random effect") of each parameter, and the correlation between them. The precise definitions of the three models being fitted are:

$g(t)=s(t)(\alpha-\beta s(t))$     "no competition"     (1)

$g(t)=s(t)(\alpha-\beta s(t)-\gamma\Phi(t))$     "competition"     (2)

$$g(t)=s(t)(\alpha-\beta s(t)-\gamma \sum_{t_0 <t'<t} \Phi(t'))  \qquad \text{"cumulative competition"} \qquad (3)$$

Residual Standard Error (RSE) summarises the difference between observed and estimated values in the model ($RSE=\sqrt{V/n}$ where V is the variance of the residuals and n is the number of observations). Aikake's An Information Criterion (AIC, Aikake (1974)) is a likelihoodbased measure with a penalisation related to the number of model parameters k: AIC = -2 ln(L) - 2k. A lower value indicates a more parsimonious model. Given the structure of the data (subsets of the complete data describe the growth curves of individual trees), the NLME approach is conceptually more appropriate, a point confirmed by the uniformly lower RSE and AIC for the NLME models. That different numbers of measurements are available for different trees (depending on their age) makes this all the more important. It transpires that there is rather large variation in growth rates, that cannot be described by a fixed set of parameters across the population. In the NLME analysis, the computed standard deviation for each parameter is on the same order as the mean, and in the case of $\gamma$, is actually larger. $\alpha$ and $\beta$ were found to be strongly correlated (in the "competition" model, $\rho_{\alpha,\beta} = 0.988$, $\rho_{\alpha,\gamma} = 0.557$, $\rho_{\beta,\gamma} = 0.481$). Despite the improved fit offered by the cumulative competition model, the basic competition model was selected for analysis and simulation due to its lack of dependence upon history (maintaining the Markov property of the process). It is also important to realise that spatiotemporal data of the type provided by these increment cores are much more laborious to collect, and as a consequence far less widely available, than the marked point process (single point in time) data that are usually used in spatial analyses.

Table 1: Estimated parameters for nonlinear growth models fitted to data from Rannoch plots 4 and 6 combined (plot 5 and 7 omitted due to missing recent management history; growth curves computed based upon increments after 1918 for individuals further than 10m from an edge). Function fitted: Gompertz with and without competition term (interaction formulated as in model description with parameters shown).

|  | **NLS** | | | **NLME** | | | |
|---|---|---|---|---|---|---|---|
|  | LS estimate | RSE | AIC | Fixed (mean) | Random (variance) | RSE | AIC |
| **Eqn. 1** | | | | | | | |
| α | 0.0426 | 0.311 | 3256.9 | 0.0931 | 0.117 | 0.132 | 8141.2 |
| β | 0.00909 | | | 0.0359 | 0.0281 | | |
| **Eqn. 2** | | | | | | | |
| α | 0.0828 | 0.269 | 1369.8 | 1.308 | 0.103 | 0.116 | 8194.0 |
| β | 0.0177 | | | 0.0318 | 0.0286 | | |
| γ | 4.46e05 | | | 6.51e05 | 6.97e05 | | |
| **Eqn. 3** | | | | | | | |
| α | 0.0684 | 0.275 | 1646.8 | 0.146 | 0.0967 | 0.115 | 8251.4 |
| β | 0.0146 | | | 0.0410 | 0.0310 | | |
| γ | 4.56e07 | | | 7.17e07 | 1.07e06 | | |

**Appendix 2 - Output comparison to Scots Pine data**

The high level of variation observed between individual growth curves in Appendix 1 suggests that the model may have difficulty in fitting real data. It is found, however, that many aspects of data from both plantation and seminatural stands are well matched by the basic model, with fixed growth parameters (Figure 1). However, allowing the asymptotic size of individuals to vary gives variation in growth trajectories, without adversely affecting other statistics (Figure 2).

Figure 1: Comparing statistics from "seminatural" datasets with simulation output. Solid line and grey envelopes in (a),(b) and (c) are simulated results. Data: Rannoch plot 4 (dashed), Rannoch 7 (fine dash), Glen Affric (dotted). Spatial correlation functions display a similar signature for all stands clustering of individuals (a), and inhibition of growth/size at short ranges (though this is not seen to a great extent in simulation, see main text) (b). Size distribution (c) varies between the stands, reflecting the management history. (d) shows the variability in size attained at a given age present in the data (individual trees represented by boxes -- data only available for Rannoch stands), compared with the simulation (×).

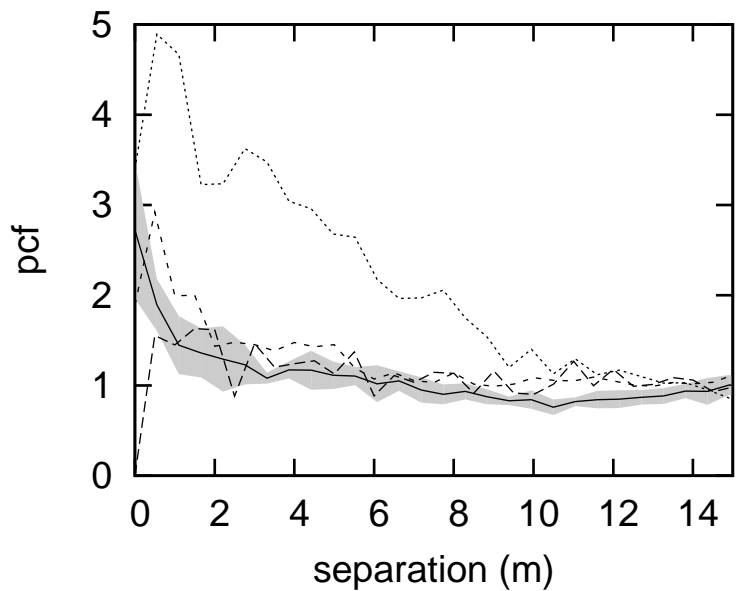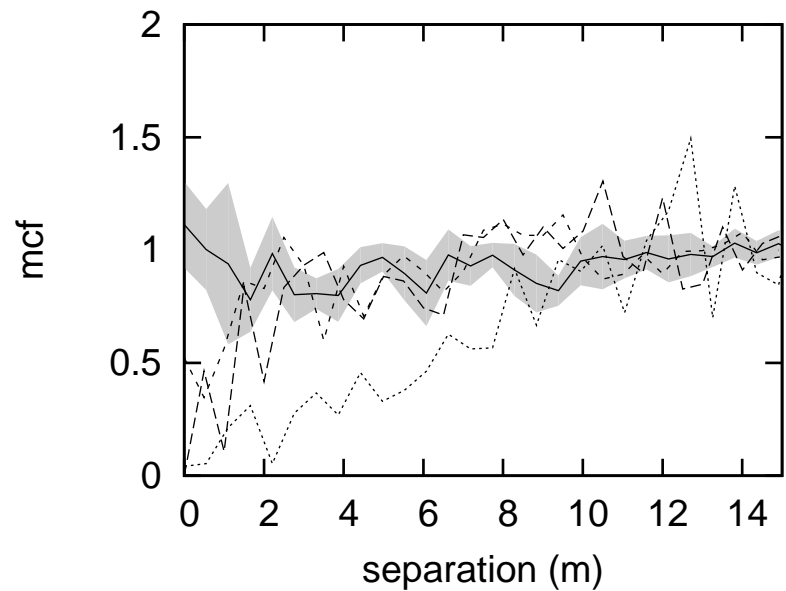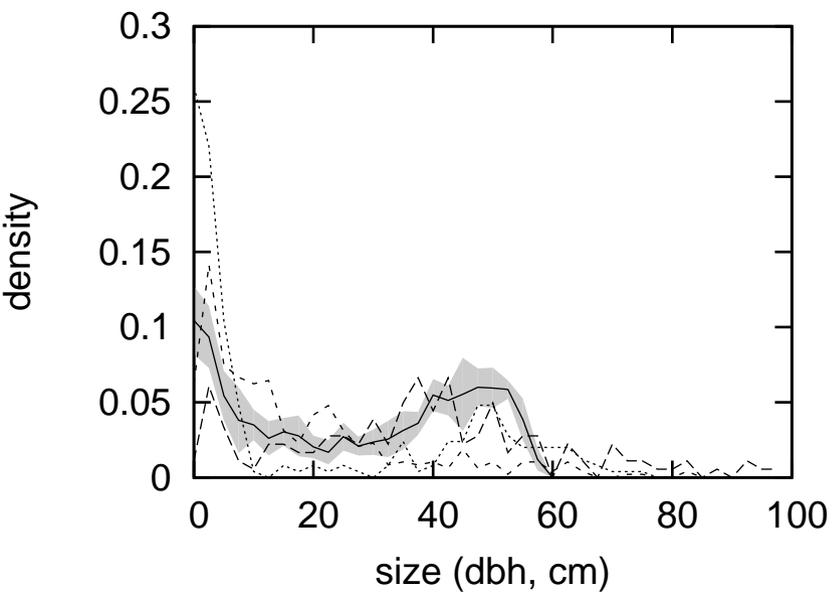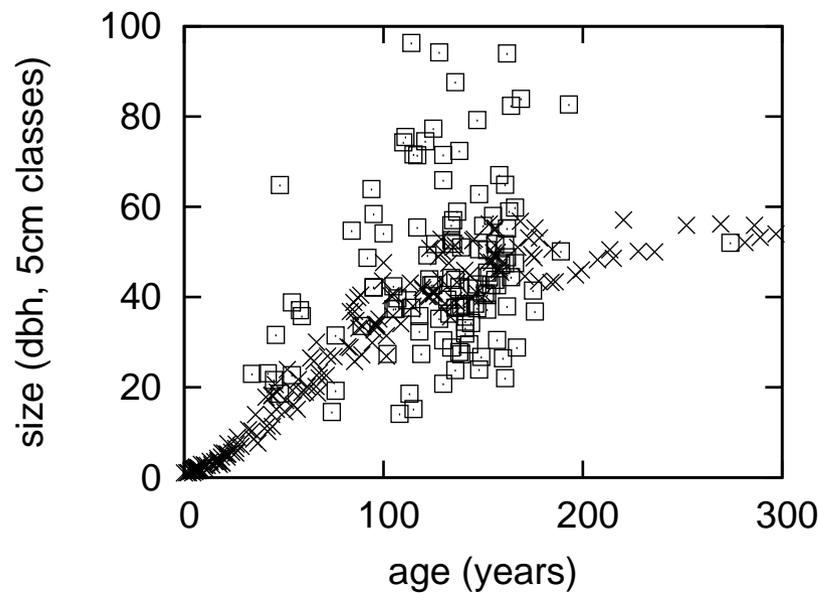

Figure 2: Results obtained from populations with fixed α, sampling exp(α /β) from observed sizes of individuals greater than 100 years old at 1990 in Rannoch plots. Again, simulation means are represented by lines within a standard deviation envelope. (a) density (dashed) and basal area (solid). Comparison with real stand data: (b) size versus age for all individuals, in the steady state, compared with Rannoch plot 4. (c) size distribution at 80 years (line) versus Glenmore plantation (6 plots) average and standard deviation (error bars). (d) size distribution at 800 years (solid line) versus Rannoch 4 (dashed line), Rannoch 7 (fine dash), and Glen Affric data (dotted).

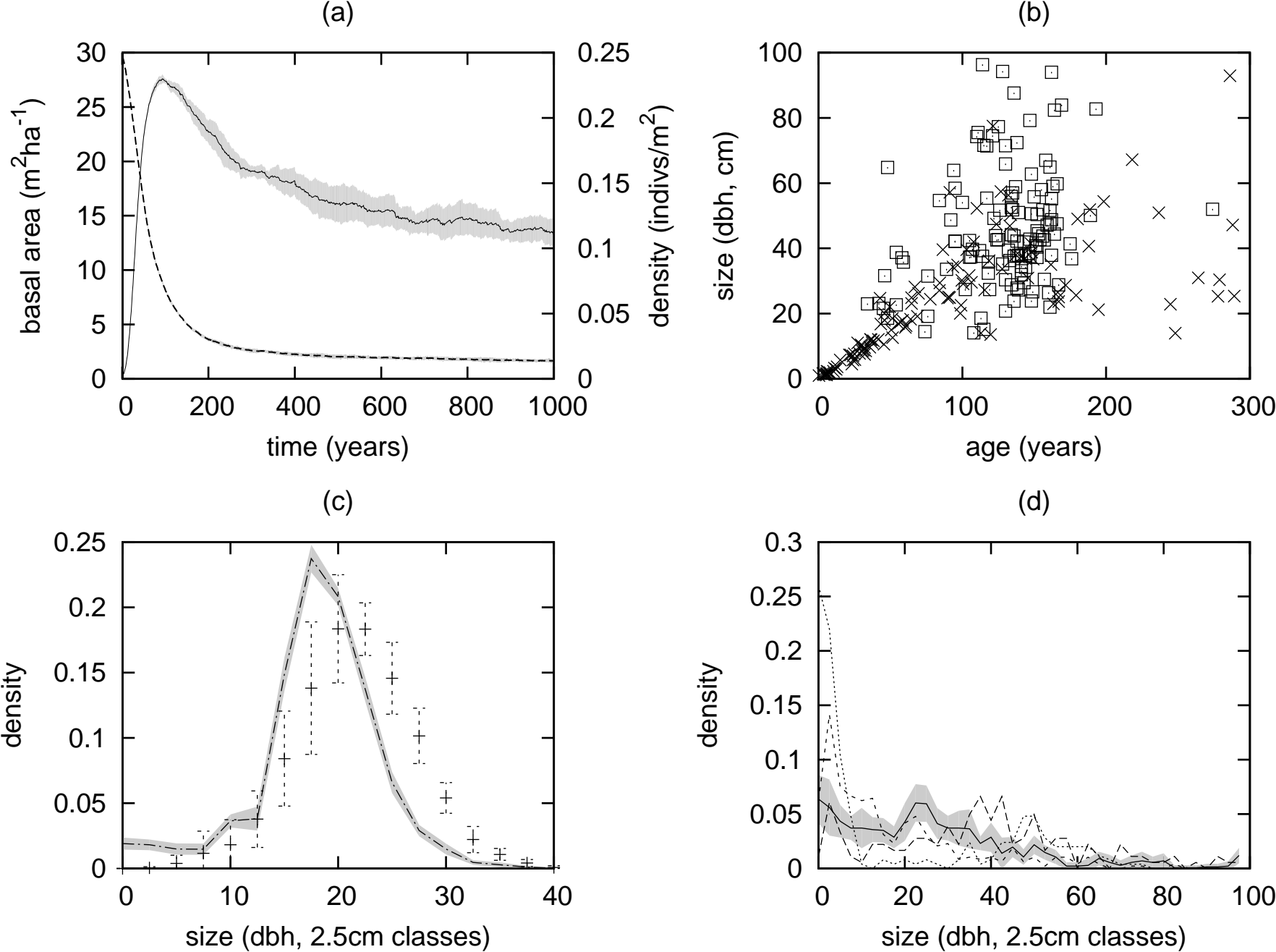

**Appendix 3 - Effects of parameter variation**

This appendix provides a brief summary of the effects of parameter variation upon various aspects of model behaviour. Model behaviour is robust: the effects described hold for at least an order of magnitude above and below the parameters used in the main text (Table 1 in main text), unless otherwise stated.

**Plantation**

The gradient of basal area increase, and the magnitude of its peak, is positively related to the speed of growth. The magnitude and time of the peak also decreases with increasing mortality rate. Fecundity has little or no effect on the transient to, or position of, the peak. Plantation size distribution is affected by a number of factors. The maximum extent is largely determined by the growth rate in the absence of competition. The shape and location of the main body of the distribution is then governed by the effect of competition on growth, and to a lesser extent, the mortality parameters. As growth interaction ($\gamma$) is increased, the mean size decreases; increasing mortality leads to an increase in the mean size (since competition is lower as a result). The variance of the distribution is affected to some extent by all parameters, but observation of the actual distributions generated indicates that the only parameter affecting the shape of the distribution of "canopy" trees is the asymmetry of competition ($k_s$: an increase widens the spread of canopy sizes). The spatial structure of the plantation is largely defined by the initial condition; any structure present at stand initiation remains evident until a very large proportion of the original trees have been removed through mortality, and juveniles have begun to replace them.

**Steady state**

In the parameter space considered (one order of magnitude above and below the parameters used in the main text, steady state density and basal area are increased by increasing fecundity or growth speed, or decreasing mortality. Decreasing mortality further leads to a decrease in steady state basal area. This somewhat surprising result occurs due to the onset of density, rather than mortality, limited individual growth (due to the resulting higher competition). This result is most likely not relevant to most temperate tree species, however, which continue growing for the duration of their lifespan. In temperate forests, multiple resource limitation, too, means that density is relatively low (this may also partly be the reason that temperate forests do not generally follow a highly optimised configuration in nature. Note that simulations with very high mortality rates were generally extinct by 800 years (the point at which the presented steady state statistics were computed), meaning that this point does not appear in the figure. The values $\rho_{closepairs}$ and $s_{closepairs}$ in Table 3 are computed by integrating over, respectively, the PCF and MCF from 0 to 5m separation. The pair density (PCF) at short ranges is insensitive to variation in most parameters, except fecundity (increasing which causes an increase), interaction mortality ($\mu_2$) and locality of interaction k d (increasing which lead to a decrease). Increasing fecundity or growth (or reducing growth interaction) increases the average size of nearby pairs (higher MCF), whilst having little or no effect on the pair density itself. Increasing k d also reduces the size of close pairs. The statistics $\rho_{canopy}$ and $s_{canopy}$ Table 3 does not show such clear patterns. $\rho_{canopy}$ is the proportion of the density greater than $0.5s_{max}$. This shows large variation with all parameters, but the pattern s are not immediately clear. It may be more instructive to consider the distributions themselves, visually. The maximum size of trees is affected most dramatically by fecundity (negative relationship) and growth parameters (increasing growth increases maximum size to a point, after which it decreases). The reason for this unexpected behaviour is likely to be that (in the case of simply increasing growth speed), more trees become larger, leading to an increase in the overall competition experienced by an individual, and a reduction in the effective asymptotic size in the steady state. Growth interaction has precisely the opposite effect.

Table 2: The effect on plantation development (as summarised by various statistics) of increasing any parameter of the model in isolation. In columns, ρ is density and s is size. Increasing "mortality" refers to increasing $\mu_1$ and $\mu_2$ whilst fixing their ratio, and increasing "growth" means increasing both $\alpha$ and $\beta$, whilst fixing their ratio.

| Parameter | Statistic $\rho_{80}$ | $BA_{80}$ | $E(s_{80})$ | $Var(s_{80})$ | $BA_{peak}$ | $t(BA_{peak})$ |
|---|---|---|---|---|---|---|
| **Rates** | | | | | | |
| f | + | + | 0 | + | 0 | + |
| mortality | - | - | + | + | - | - |
| growth | - | + | + | + | + | - |
| **interaction** | | | | | | |
| $\mu_2$ | - | - | + | 0 | - | - |
| $\gamma$ | + | - | - | - | - | + |
| **kernels** | | | | | | |
| $k_d$ | + | 0 | + | + | + | 0 |
| $k_s$ | 0 | 0 | 0 | + | 0 | 0 |

Table 3: The effect on steady state behaviour (as summarised by various statistics) of increasing any parameter of the model in isolation. Again, $\rho$ is density and s is size. Increasing "mortality" refers to increasing $\mu_1$ and $\mu_2$ whilst fixing their ratio, and increasing "growth" means increasing both $\alpha$ and $\beta$, whilst fixing their ratio. Canopy density is relative to total density (proportion of individuals > 50% of maximum size). A star indicates variation across the observed parameter range, but no clear trend.

| | Statistic | | | | | |
|---|---|---|---|---|---|---|
| Parameter | $\rho$ | BA | $\rho_{canopy}$ | $s_{canopy}$ | $\rho_{closepairs}$ | $s_{closepairs}$ |
| **Rates** | | | | | | |
| f | + | + | - | - | + | + |
| mortality | - | - | + | + | 0 | - |
| growth | + | + | - | 0* | 0 | + |
| **interaction** | | | | | | |
| $\mu_2$ | - | - | + | + | - | - |
| $\gamma$ | - | - | - | 0* | 0 | - |
| **kernels** | | | | | | |
| $k_d$ | + | + | 0 | 0 | - | - |
| $k_s$ | 0 | 0 | 0* | 0 | 0 | 0* |